\documentclass[10pt, conference, letterpaper]{IEEEtran}
\usepackage{xcolor}
\IEEEoverridecommandlockouts
\usepackage{amsmath,amssymb,amsfonts}
\usepackage{gensymb}
\usepackage{algorithmic}
\usepackage{graphicx}
\usepackage{textcomp}
\usepackage{xcolor}
\usepackage{xspace}
\usepackage{enumitem}
\usepackage{url}
\usepackage{caption}
\usepackage{subcaption}
\usepackage{multirow}
\usepackage[normalem]{ulem}
\def\BibTeX{{\rm B\kern-.05em{\sc i\kern-.025em b}\kern-.08em
    T\kern-.1667em\lower.7ex\hbox{E}\kern-.125emX}}
\graphicspath{ {./images/} }

\newcommand{\ie}{\textit{i.e.,}\xspace}
\newcommand{\eg}{\textit{e.g.,}\xspace}

\begin{document}

\title{Impact of Device Thermal Performance on 5G mmWave Communication Systems  \thanks{\hrule \vspace{4pt} This research  on is supported in part by the National Science Foundation under grant numbers CNS-1618692 and CNS-1618836.}}

\author{\IEEEauthorblockN{
Muhammad Iqbal Rochman\IEEEauthorrefmark{1},
Damian Fernandez\IEEEauthorrefmark{2},
Norlen Nunez\IEEEauthorrefmark{2},
Vanlin Sathya\IEEEauthorrefmark{1},\\
Ahmed S. Ibrahim\IEEEauthorrefmark{2}, 
Monisha Ghosh\IEEEauthorrefmark{1}, and 
William Payne\IEEEauthorrefmark{1}}
\IEEEauthorblockA{
\IEEEauthorrefmark{1}Department of Computer Science, University of Chicago, Chicago, Illinois, USA\\
\IEEEauthorrefmark{2}Department of Electrical and Computer Engineering, Florida International University, Miami, Florida, USA\\ 
Email: \{muhiqbalcr,vanlin,monisha,billpayne\}@uchicago.edu, \{dfern265,nnune047,aibrahim\}@fiu.edu}}


\maketitle
\thispagestyle{plain}
\pagestyle{plain}

\begin{abstract}


5G millimeter wave (mmWave) cellular networks have been reported to deliver 1-2 Gbps downlink throughput, via speed-tests. However, these speed-tests capture only a few seconds of throughput and are not representative of sustained throughput over several minutes. We report the first measurements of sustained throughput in three cities, Miami, Chicago, and San Francisco, where we observe throughput throttling due to rising skin temperature of the phone when it is connected to a deployed 5G mmWave base-station (BS). Radio Resource Control (RRC) messaging between the phone and the BS indicates the reduction in the number of aggregated mmWave channels from 4 to 1 followed by a switch to 4G. We corroborate these measurements with infra-red images as the phone heats up. Thus, mmWave throughput will be limited not by network characteristics but by device thermal management.

\begin{IEEEkeywords}
5G, mmWave, throughput, thermal, throttling, skin temperature, CPU temperature.
\end{IEEEkeywords}


\end{abstract}


\section{Introduction}\label{sec:intro}

5G New Radio (5G NR) cellular networks are being rapidly deployed around the world in low ($<1$~GHz), mid ($1-6$~GHz), and high ($>24$~GHz) bands. Of these, the high, or mmWave bands, are being deployed predominantly in dense urban areas in the US while the low and mid-bands, including the C-Band (3.7 - 3.98 GHz) are witnessing deployments that can provide wider coverage in suburban and rural areas due to the favorable propagation characteristics.
Recent measurements on deployed 5G mmWave networks in major US cities demonstrate that indeed 5G mmWave can deliver extremely high throughput in the range of 1 - 2 Gbps~\cite{narayanan2020lumos5g,narayanan2020first,rochman2022comparison}. These high throughputs are enabled by aggregating up to eight 100 MHz mmWave channels, depending on network and device capabilities.

However there is still a number of challenges associated with guaranteeing QoS in 5G mmWave: beam-tracking, beam management, building blockage, and rain attenuation, to name a few, and these are being addressed in many research efforts ~\cite{giordani2018tutorial,li2020beam}. In this paper we address a question that has received less attention by the research community: what is the sustained downlink throughput, over several minutes, that can be delivered by a 5G mmWave connection? Most reported speed-test measurements use commonly used speedtest apps such as Ookla\footnote{\url{https://www.speedtest.net/}} or the FCC Speedtest\footnote{\url{https://play.google.com/store/apps/details?id=com.samknows.fcc}}, where the test runs for only 5 - 10 seconds and is not indicative of the average throughput when an application is running over several minutes at the high throughput.
We postulate that a high-throughput data transfer over several minutes using multiple mmWave channels will cause device heating with a resultant increase in the skin temperature that will then trigger throttling of the throughput until the device cools to acceptable levels.
We present results from detailed experiments conducted with consumer 5G smartphones operating over deployed 5G mmWave networks to demonstrate that indeed this phenomenon occurs repeatedly when the ambient temperatures are high enough to deter the cooling of the device. 
We demonstrate that as the skin temperature measured by the device increases, the number of mmWave channels being aggregated drops from 4 to 1 followed by handover to 4G LTE, with the throughput dropping at each step.
With external cooling, \eg using an ice-pack or low ambient temperature (\eg on a Chicago winter day), high throughput can be sustained over several minutes. Furthermore, we identify explicit message exchanges in the Radio Resource Control (RRC) layer between the user equipment (UE) and the base-station (BS) that confirm that the reason for handing over to 4G LTE is thermal and not network congestion or other considerations.
Lastly, we used an infra-red (IR) camera to further corroborate the effect of temperature rise at the mmWave antenna locations on throughput.


\begin{table}[!t]
\caption{Experiment Parameters}
\centering
\begin{tabular}{|l|c|c|c|}
\hline
\bfseries Parameter & \bfseries Value \\
\hline
Operator & Verizon (Band n261/28 GHz) \\
\hline
\# of experiment locations & Chi.: 2, Mia.: 1, SF: 1\\
\hline
Device model & Google Pixel 5 \\
\hline
\# of devices used & Chi.: 2, Mia.: 1, SF: 1 \\
\hline
Cumulative \# of meas. runs & 32 \\
\hline
Distance between BS and UE & $\sim$2 meter \\
\hline
Average RSRP over all meas. & -92.63 dBm \\
\hline

\end{tabular}
\label{tab:parameters}
\vspace{-0.5cm}
\end{table}

\begin{figure*}[t!]
\centering
\begin{subfigure}{.43\textwidth}
  \includegraphics[height=6.3cm]{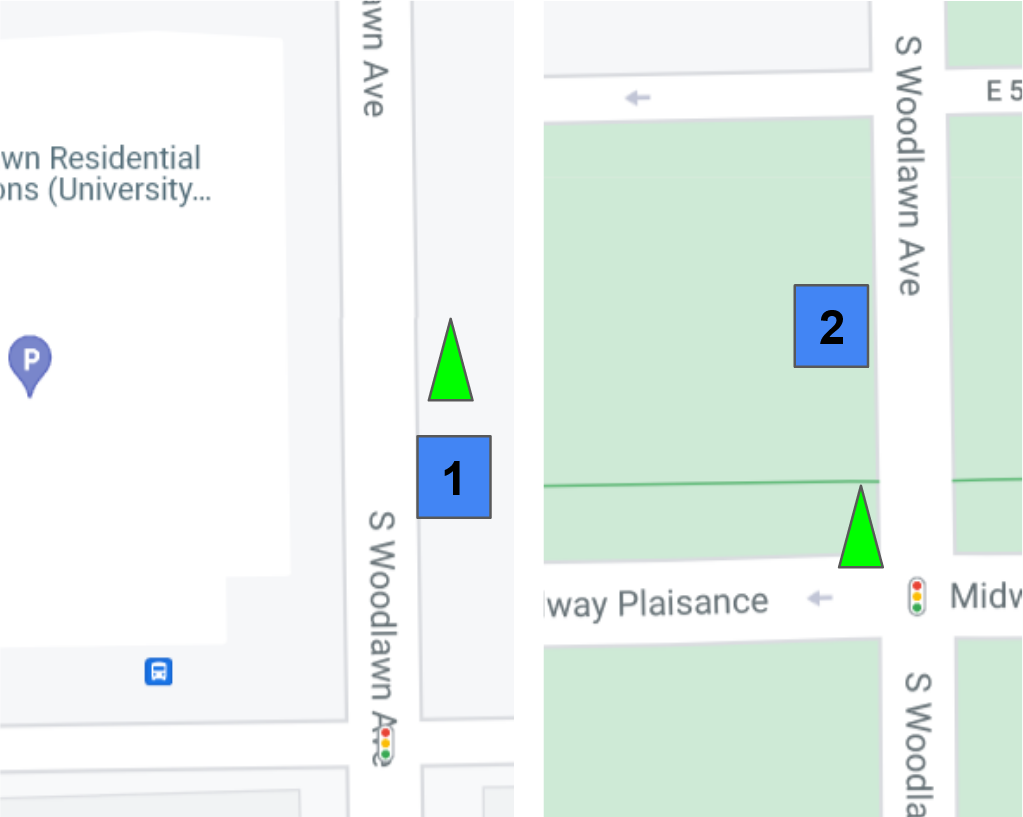}
  \caption{Chicago}\label{fig:Chicago_map}
\end{subfigure}
\hfill
\begin{subfigure}{.26\textwidth}
 \includegraphics[height=6.3cm]{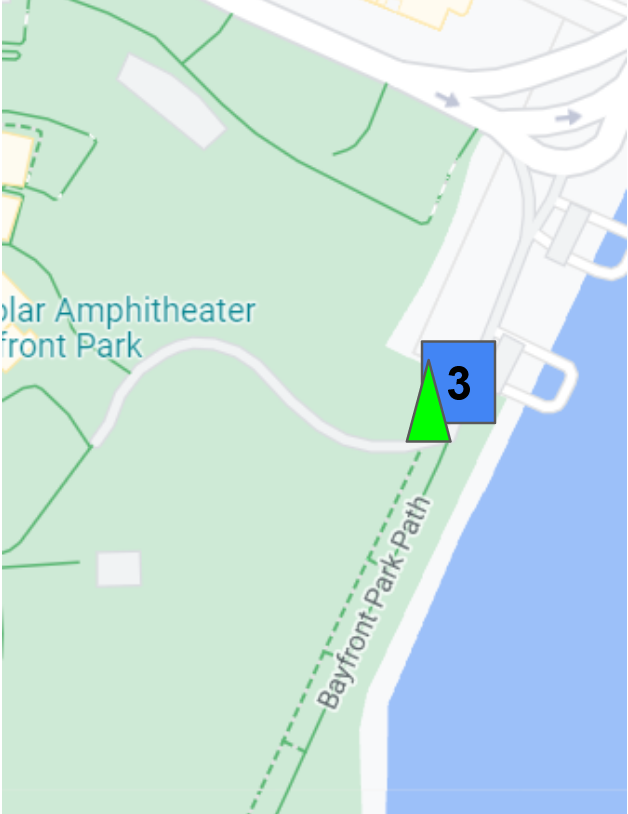}
    \caption{Miami}\label{fig:Miami_map}
\end{subfigure}
\hfill
\begin{subfigure}{.25\textwidth}
  \includegraphics[height=6.3cm]{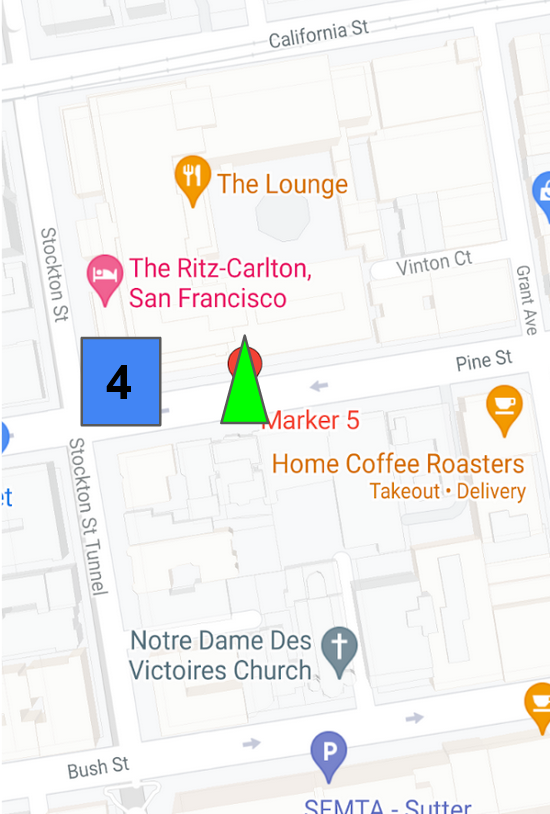}
  \caption{San Francisco}\label{fig:SF_map}
\end{subfigure}
\caption{Measurements locations in the three different cities (see Table \ref{tab:locations}).}
\vspace{-0.3cm}
\label{fig:maps}
\end{figure*}

\begin{table*}[!t]
\caption{Experiment Locations}
\centering
\begin{tabular}{|l|c|c|c|c|c|c|}
\hline
\bfseries Location & \bfseries \# & \bfseries Address & \bfseries GPS coordinate & \bfseries Traffic used & \bfseries \# of meas. runs & \bfseries Ambient temp. at meas. time\\
\hline
\multirow{2}{*}{\bfseries Chicago} 
& 1 & 61st St. \& Woodlawn Ave. & 41.7874024,-87.5965241 & BG DL+FCC ST & 16 & \multirow{2}{*}{Oct: $\sim$$24\degree$~C, Jan: $\sim$$-10\degree$~C}\\
\cline{2-6}
& 2 & 61st St. \& Midway Plai. & 41.7844559,-87.5962098 & BG DL+FCC ST & 9 & \\
\hline
\bfseries Miami & 3 & Bayfront Park & 25.7753436,-80.1853549 & BG DL & 6 & Sep: $\sim$$31\degree$~C \\
\hline
\bfseries San Francisco & 4 & Pine \& Stockton St. & 37.7914117,-122.407218 & BG DL+FCC ST & 1 & Sep: $\sim$$15\degree$~C\\
\hline
\end{tabular}
\label{tab:locations}
\vspace{-0.5cm}
\end{table*}

\section{Related work}
\label{sec:background}
Recent literature~\cite{narayanan2020lumos5g,narayanan2020first} has demonstrated the feasibility of achieving very high throughput with consumer smartphones over commercially deployed 5G mmWave, in spite of the well-known limitations of mmWave propagation due to beam tracking, beam management, mobility management and building blockage. Advanced techniques, based on machine learning and artificial intelligence, have been proposed for addressing these limitations, for example in ~\cite{giordani2018tutorial,li2020beam}.  Most recently, \cite{narayanan2022comparative} presents detailed measurements of 5G mmWave deployments by two major commercial 5G operators in the US in diverse environments
using smartphone-based tools.
The measurement-driven propagation analysis demonstrated performance differences due to terrain, frequency of operation, antenna pattern, etc. However, the relationship between device temperature and sustained 5G mmWave throughput was not explored. The results presented in this paper seek to address the gap in the literature on the effect of device thermal management on end-user throughput on 5G mmWave networks.

In particular, we seek to demonstrate that the drop in throughput is indeed due to thermal. According to the 3GPP standard~\cite{3gpp20205grrc}, a UE can provide information to the BS about its thermal state via the the RRC\_CONNECTED message field. Upon receiving such a message from the UE, the BS will respond by temporarily reducing the number of aggregated data streams, in both component carriers and MIMO layers, in both downlink and uplink transmissions until the thermal warning messages are no longer received. This reduction in component carriers (e.g. reduction from 4 to 1 mmWave channels) will lead to a reduction in throughput until the skin temperature drops to below a pre-specified threshold.

\section{Measurement Tools and Methodology}
\label{sec:methodology}

In order to demonstrate the effect of device skin temperature on sustained throughput over 5G mmWave, the following requirements need to be met:
\begin{itemize}
\item A sustained download of a high-bandwidth data stream over $\sim$15 minutes while connected to a 5G mmWave BS.
\item A method of measuring temperature while the download is occurring, and
\item A method of extracting RRC messaging between the UE and the BS while the download is occurring.
\end{itemize}

In this section, the tools and methodology used in this paper to satisfy the above requirements are described.

Table~\ref{tab:parameters} summarizes the parameters of the experiments conducted in two locations in Chicago, one location in Miami and one location in San Francisco. Data was collected in all 3 locations over September - October 2021, and in Chicago in January 2022 for performance comparison under cooler ambient temperatures.  
All experiments were conducted using the same UE model and network: Google Pixel 5, running Android 11 on a Verizon network with an unlimited data plan\footnote{Subscribed Verizon plan indicates a throttling after 50 GBytes for 4G and 5G low/mid-band data, and no throttling for 5G mmWave data.}. Fig.~\ref{fig:maps} shows the specific measurement locations in Chicago, Miami and San Francisco, while Table~\ref{tab:locations} shows detailed information of each location. The Verizon 5G mmWave network at each location utilizes band n261 at $28$~GHz.

Downlink throughput saturation is achieved using a combination of two methods:
\begin{itemize}
    \item \textbf{Background Download (BG DL)} using HTTP download of a 10 GB dataset file~\cite{YUV}.
    \item \textbf{FCC Speed Test (FCC ST)} app: the 5 sec downlink throughput test is run repeatedly to ensure that the link stays saturated continuously.
\end{itemize}
Thermal throttling was observed using either one of the above methods, but combining both methods ensures that the link is fully saturated. The Miami measurements used only the BG DL traffic, while BG DL + FCC ST was used in the Chicago and San Francisco measurements. Due to this minor difference in methodology, there are two separate throughput measurements: \textbf{PHY} level throughput collected by Network Signal Guru (NSG), described below, and \textbf{APP} level throughput collected by FCC ST. Using all the measurements reported in this paper, we verified that, as expected, the APP throughput is always lower than the PHY throughput. APP throughput values are easier to extract from FCC ST than PHY throughput from NSG (requiring manual data input). Thus, these different types of throughput measurements are carefully separated and only the same type of throughput values are compared whenever needed in our analysis.

The following Android apps were used to collect measurements systematically for the experiments described in \S\ref{sec:results}:
\begin{itemize}
\item \textbf{SigCap}~\cite{rochman2022comparison}, an Android app developed at the University of Chicago which collects Global Positioning System (GPS), time and location information along with signal and network parameters (\eg 4G and 5G RSRP, RSRQ, RSSI, PCI, 4G frequency, etc) through APIs that extract information directly from the modem chip and hence is compliant to relevant standards. Instantaneous skin, CPU and GPU temperature measurements from the APIs were added to SigCap~\cite{androidAPI_temperature} for the work reported in this paper.
\item \textbf{Network Signal Guru (NSG)}\footnote{\url{https://m.qtrun.com/en/product.html}}, a commercial app that utilizes the phone's root capability to provide more detailed information about the transmission such as operating frequency, number of carrier components, bandwidth, PHY throughput, and RRC messaging. However, data export from NSG is difficult compared to SigCap and hence we use it only for certain measurements.
\end{itemize}

\section{Experimental Results}\label{sec:results}

In this section, we demonstrate the impact of 5G mmWave transmissions on the device temperature as measured by SigCap, and the resulting effect on downlink throughput under various operating conditions that impact the UE temperature (\eg~phone cover and ambient temperature).
Furthermore, we confirm our findings using an IR camera to image the phone as the mmWave transmission progressed over time to demonstrate that the throughput drop correlated with the rise in temperature at the mmWave antenna locations.

\subsection{Impact of 5G mmWave on UE Temperature}

\begin{figure}[!t]
    \centering
    \begin{subfigure}{0.5\textwidth}
        \includegraphics[width=1\linewidth]{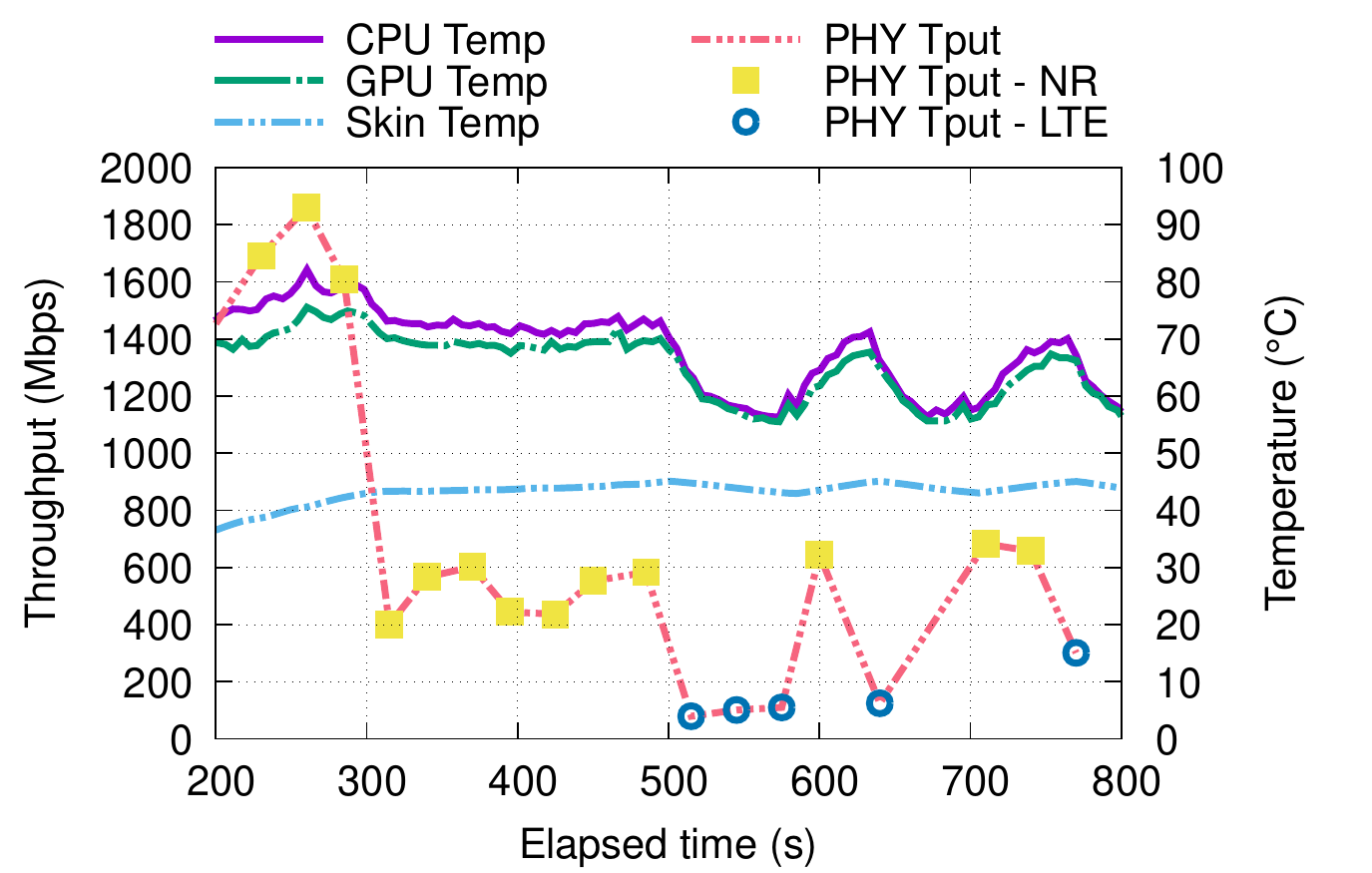}
        \caption{Throughput degradation correlated with device temperatures.}
        \label{fig:representative_tput_temp}    
    \end{subfigure}
    \hfill
    \begin{subfigure}{0.5\textwidth}
        \includegraphics[width=1\linewidth]{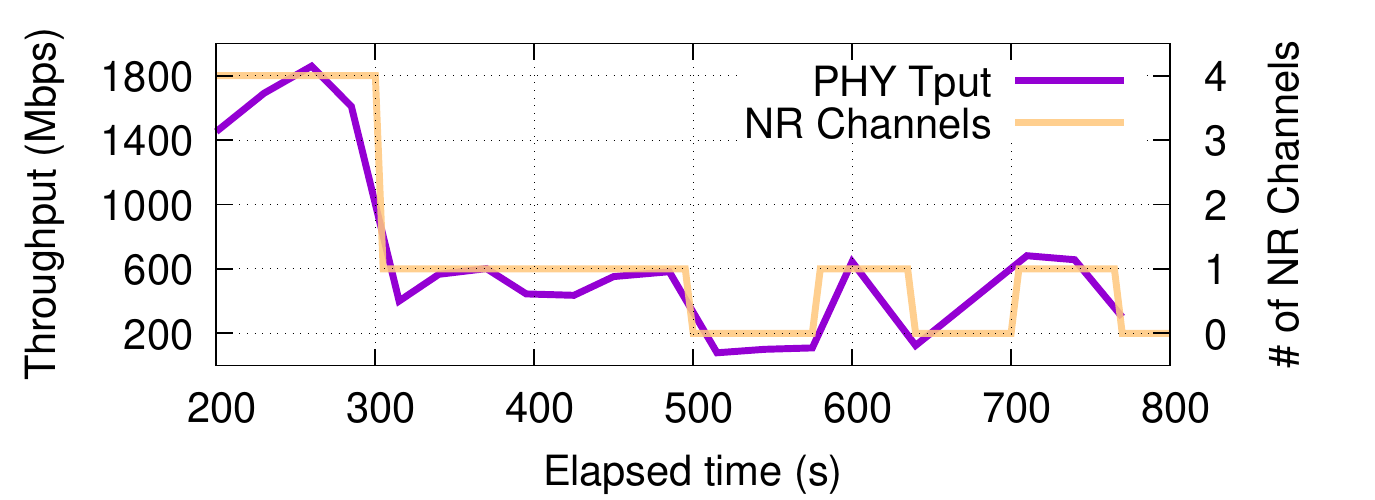}
        \caption{Correlation of throughput with number of NR channels.}
        \label{fig:representative_tput_nrch}    
    \end{subfigure}
    \caption{Representative results of throughput degradation.}
    \label{fig:representative_tput}
\vspace{-0.2in}
\end{figure}

\noindent\textbf{5G mmWave throughput and UE temperature Vs. time} In order to demonstrate the effect of 5G mmWave on device temperature we conducted numerous measurements using the tools and methodologies describe in the previous section. We performed a total of 32 measurement runs over all locations, where each measurement run starts with a cool phone. Fig.~\ref{fig:representative_tput} shows a representative measurement at Location 1 in Chicago (taken on Oct 9, 2021) using the combined BG DL + FCC ST method. The PHY data rate and number of mmWave channels are manually transcribed from NSG, while the temperature data is collected from SigCap, both taken at $5$~sec intervals and synchronized using timestamps from both apps. Fig.~\ref{fig:representative_tput} only shows the PHY throughput when the FCC ST is running (at $\sim$20~sec intervals) in order to display the conditions when the downlink to the UE is fully loaded.

Fig.~\ref{fig:representative_tput_temp} shows that a PHY throughput of almost $2$~Gbps was achieved soon after the experiment was started at the $200$~sec mark, which is the result of aggregating 4 mmWave channels as shown in Fig.~\ref{fig:representative_tput_nrch}. The throughput increase is accompanied by a rise on all three temperature measurements: skin, CPU and GPU.
At the $300$~sec mark, the number of aggregated mmWave channels reduces to 1 and the resultant throughput is reduced significantly. At this point, the CPU and GPU temperatures are reduced slightly, but the skin temperature does not reduce sufficiently to restore the throughput to the levels seen at the beginning of the experiment. The download was completed at $800$~sec.


\begin{figure}[!t]
    \centering
    \includegraphics[width=1\linewidth]{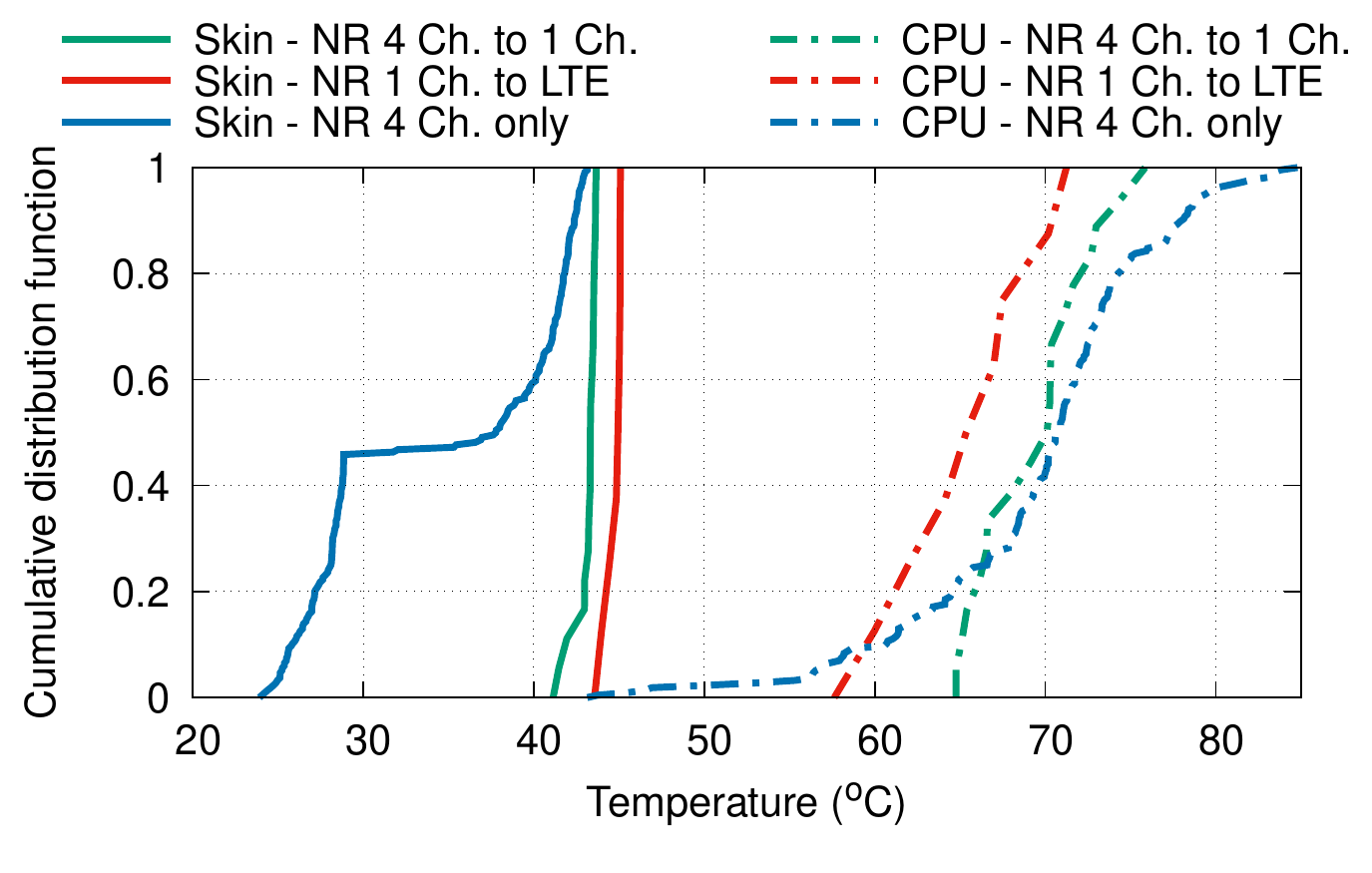}
    \caption{CDF of CPU and skin temperature when number of NR channels changed, and when NR channel is 4.}
    \label{fig:cdfTrigTemp}
    \vspace{-0.2in}
\end{figure}

\vspace{0.2em}\noindent\textbf{Analysis of skin temperature effect on throughput}. We observe two events: \textbf{(1)} when the number of 5G mmWave channels is reduced from 4 to 1 (\ie~$300$~sec on Fig.~\ref{fig:representative_tput_temp}), and \textbf{(2)} when the device is handed over to the LTE network ($500$~sec). At both events, we recorded a "\textit{Secondary Cell Group Failure}" signalling packet in the NSG log, which shows compliance to the 3GPP standard~\cite{3gpp20205grrc}. Moreover, using the Android Temperature API~\cite{androidAPI_temperature}, we obtained the static temperature threshold values: $96\degree$~C for CPU and GPU, and $43\degree$~C for Skin. Fig.~\ref{fig:cdfTrigTemp} shows the skin and CPU temperature distribution of all our data from all locations for the following cases: (i) the temperature when 4 mmWave channels are being aggregated, (ii) the temperature  just after the switch from 4 mmWave channels to 1 mmWave channel, and (ii) the temperature just after the switch from 5G mmWave to LTE.  We omit GPU temperature since we observe that the CPU and GPU temperatures are similar. The figure clearly shows that throttling to 1 mmWave channel happens mostly at skin temperature of $\sim$$43\degree$~C, while throttling down to LTE happens mostly at skin temperature of $\sim$$45\degree$~C.
On the other hand, the CPU temperature does not exhibit any correlation with the events since the its threshold is never crossed. Hence, we infer that the skin temperature is the trigger that causes the throughput degradation. While it appears that there is some oscillation between states, on a larger time scale, we can still observe an on-off pattern as shown by Fig.~\ref{fig:representative_tput} between the $500$ to $770$~sec mark.

\begin{figure}[!t]
    \centering
    \includegraphics[width=1\linewidth]{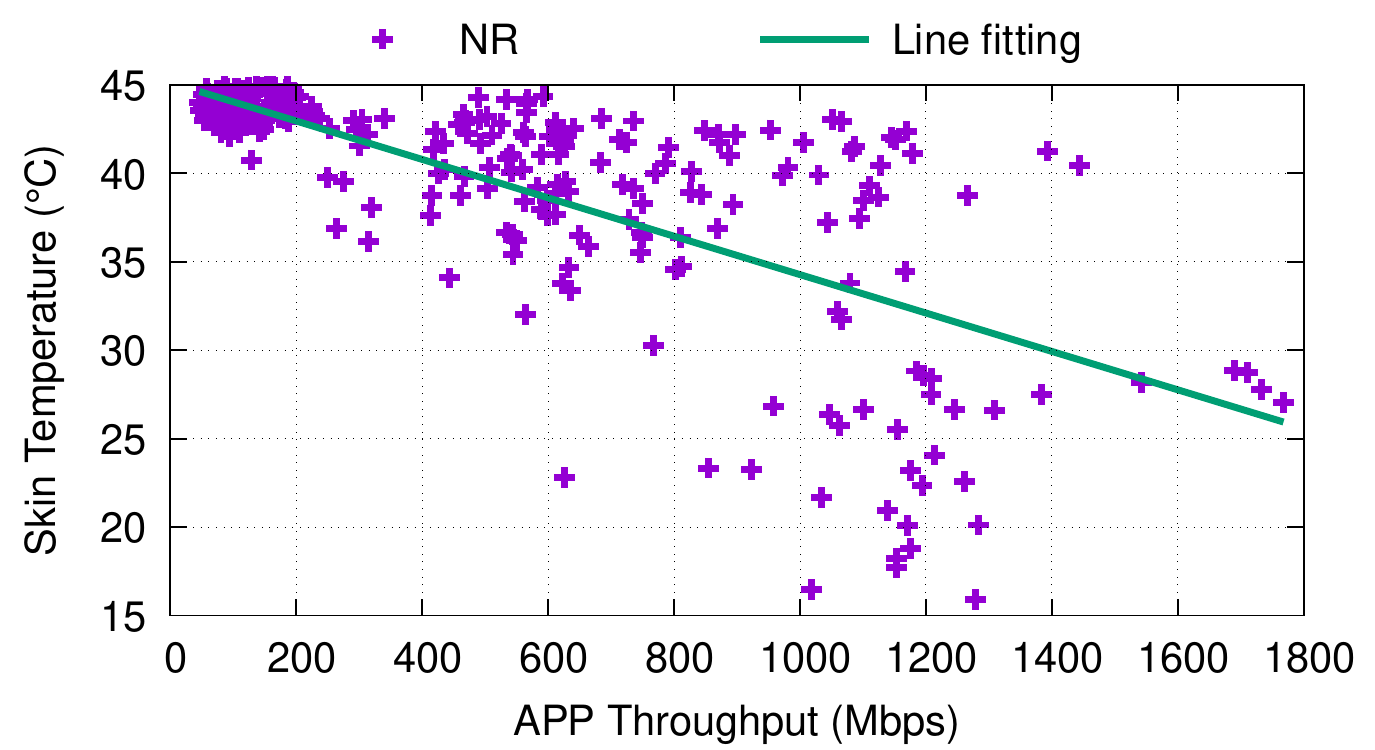}
    \caption{Correlation between APP throughput and Skin temperature.}
    \label{fig:tpt_temp}
    \vspace{-0.2in}
 \end{figure}
 
Fig.~\ref{fig:tpt_temp} displays all 5G mmWave measurements collected at Chicago and San Francisco, using both BG DL and FCC ST to saturate the downlink transfer. Clearly, the  higher skin temperature correlates to lower 5G mmWave throughput, with lower throughput values recorded mostly in summer (Sep-Oct) and the higher values recorded in Chicago in winter (Jan).

\subsection{Thermal performance as a function of ambient conditions}


\begin{figure}[!t]
    \centering
    \begin{subfigure}{0.5\textwidth}
        \includegraphics[width=1\linewidth]{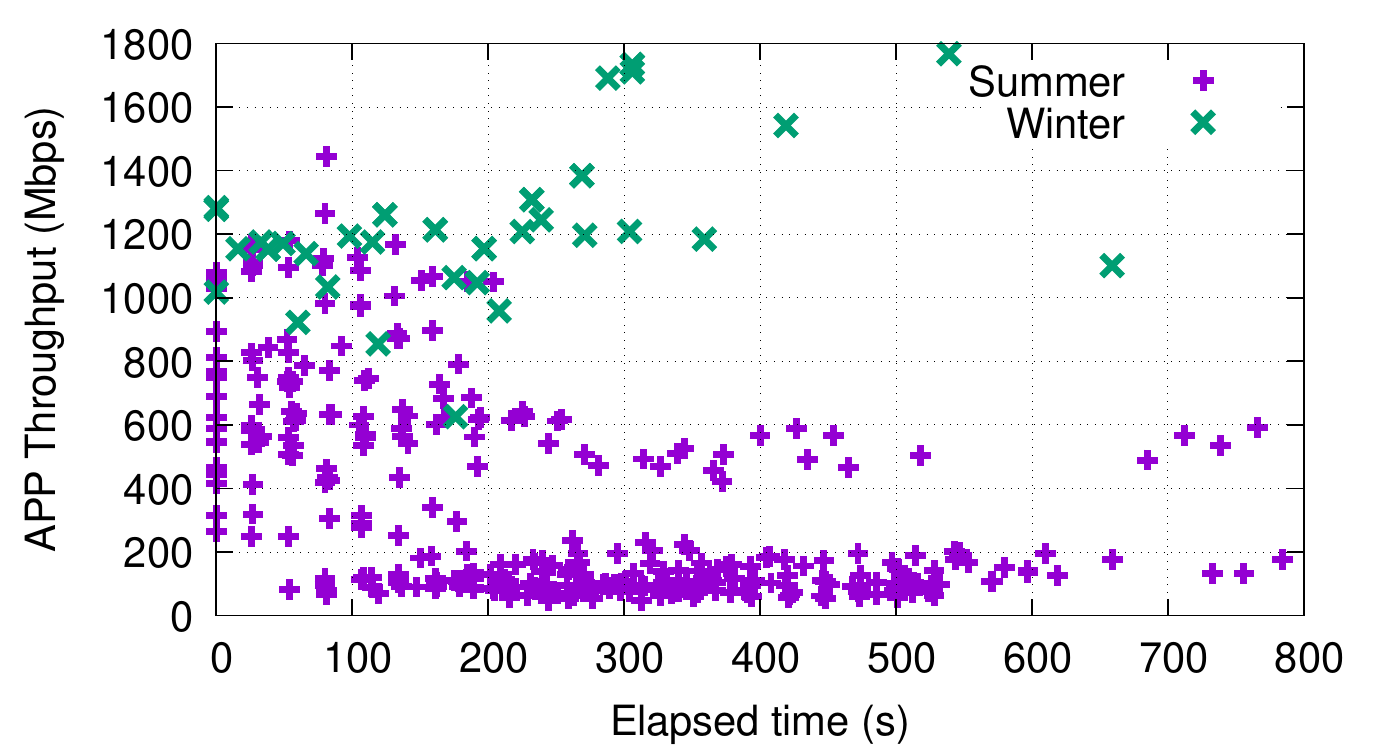}
        \caption{APP Throughput comparison between summer and winter in Chicago.}
        \label{fig:time_tput_season}
    \end{subfigure}
    \begin{subfigure}{0.5\textwidth}
        \includegraphics[width=1\linewidth]{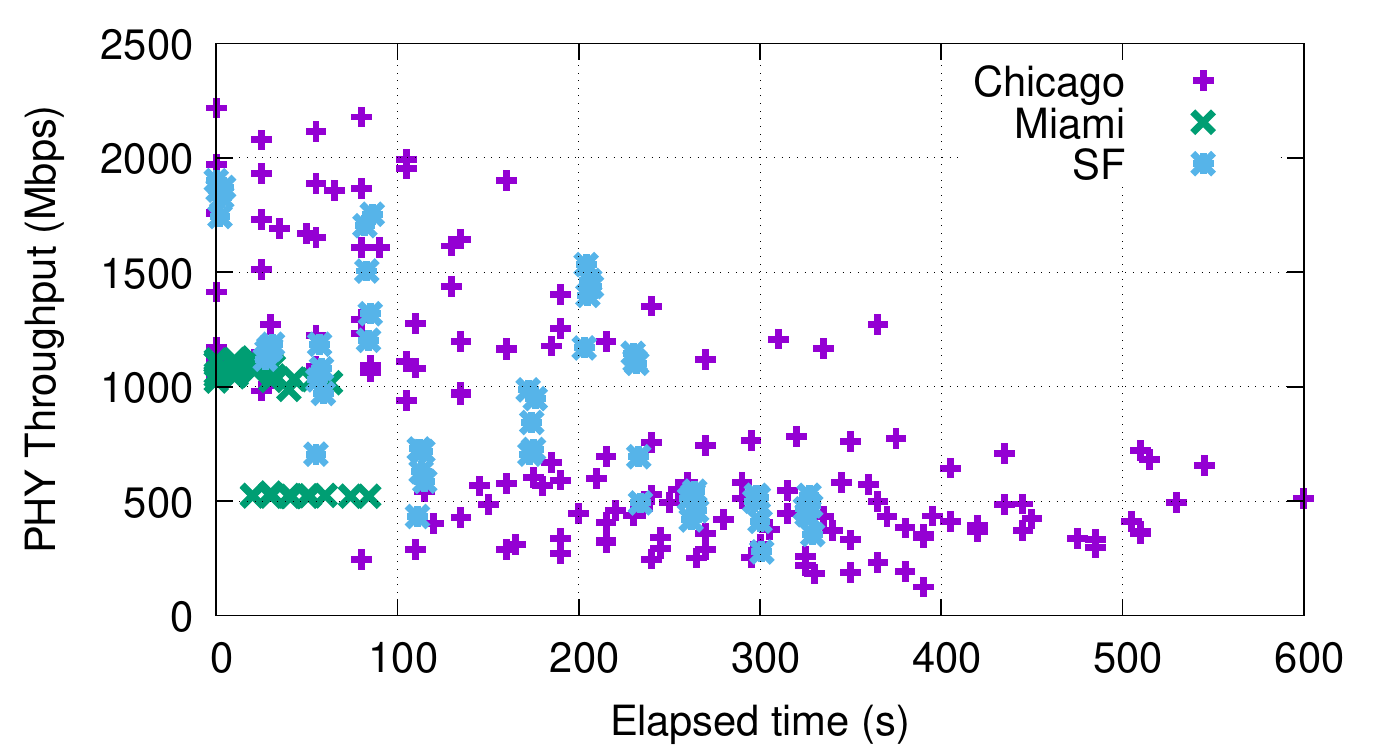}
        \caption{PHY Throughput comparison between Chicago, Miami, and San Francisco in summer.}
        \label{fig:time_tput_city}
    \end{subfigure}
    \caption{Throughput over multiple experiments, time-normalized to the first data}
    \label{fig:time_normtime}
    \vspace{-0.2in}
\end{figure}

\noindent\textbf{Effect of ambient temperature over seasons and location.} Fig.~\ref{fig:time_normtime} shows the mmWave throughput versus time, where the time axis has been normalized \ie~$0$~sec is the timestamp of the first data point. Fig.~\ref{fig:time_tput_season} shows the comparison of APP throughput between summer (Sep-Oct) and winter (Jan) in Chicago. These measurements are from Location 1 and 2, using the combination of FCC ST and BG DL. It is clear from the figure that in the warmer months, when the ambient temperature was $\sim$$24\degree$~C there is a degradation of throughput after $200$~sec, while no such degradation is observed in the winter months when the ambient temperature was $\sim$$-10\degree$~C.

Fig.~\ref{fig:time_tput_city} shows the comparison of measurements in Chicago, Miami, and San Francisco collected in summer. Since the data in Miami was captured using BG DL traffic only, PHY throughput from NSG is used in this analysis. The throughput in Miami data degrades faster (at $\sim$$60$~sec) than Chicago and San Francisco, which can be explained by the climate difference between these cities and the time of experiment. The Miami data was taken with ambient temperature of $\sim$$31\degree$~C, while Chicago and San Francico data was taken with ambient temperature of $\sim$$24\degree$~C and $\sim$$15\degree$~C, respectively.

\begin{figure}[!t]
    \centering
    \includegraphics[width=1\linewidth]{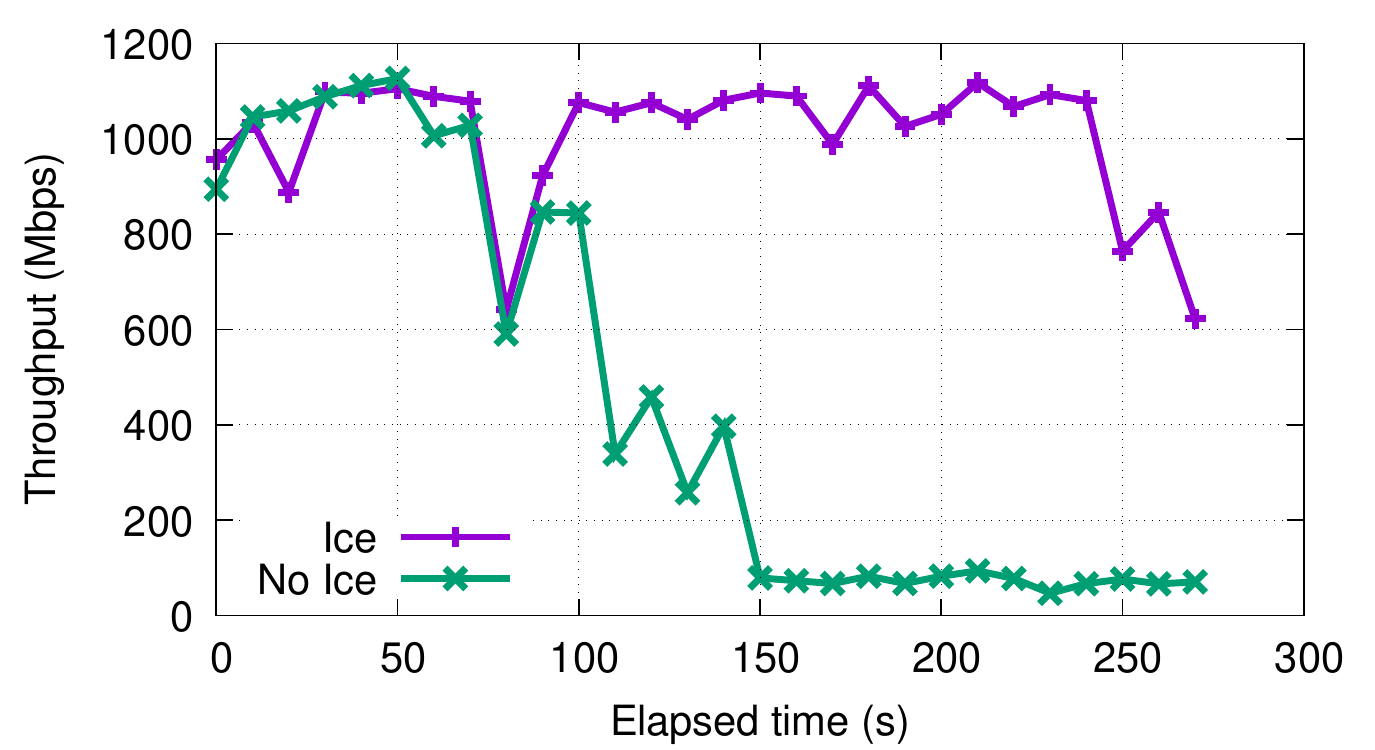}
    \caption{Prolonging high rate of 5G NR session via an ice bag.}
    \label{fig:ice_Tpt_Temp}
\vspace{-0.5cm}
\end{figure}

\vspace{0.2em}\noindent\textbf{Effect of external cooling and phone cover.}
To further confirm the correlation between skin temperature and reduced 5G mmWave throughput, the following experiment was conducted in Miami in summer. Measurements were taken with the phone either held in the hand or placed on an ice-pack.
Fig.~\ref{fig:ice_Tpt_Temp} shows that a mmWave throughput above $1$~Gbps is sustained when the phone is placed on top of an ice-pack. 





\begin{figure}[!t]
    \centering
    \begin{subfigure}{\linewidth}
        \includegraphics[width=1\linewidth]{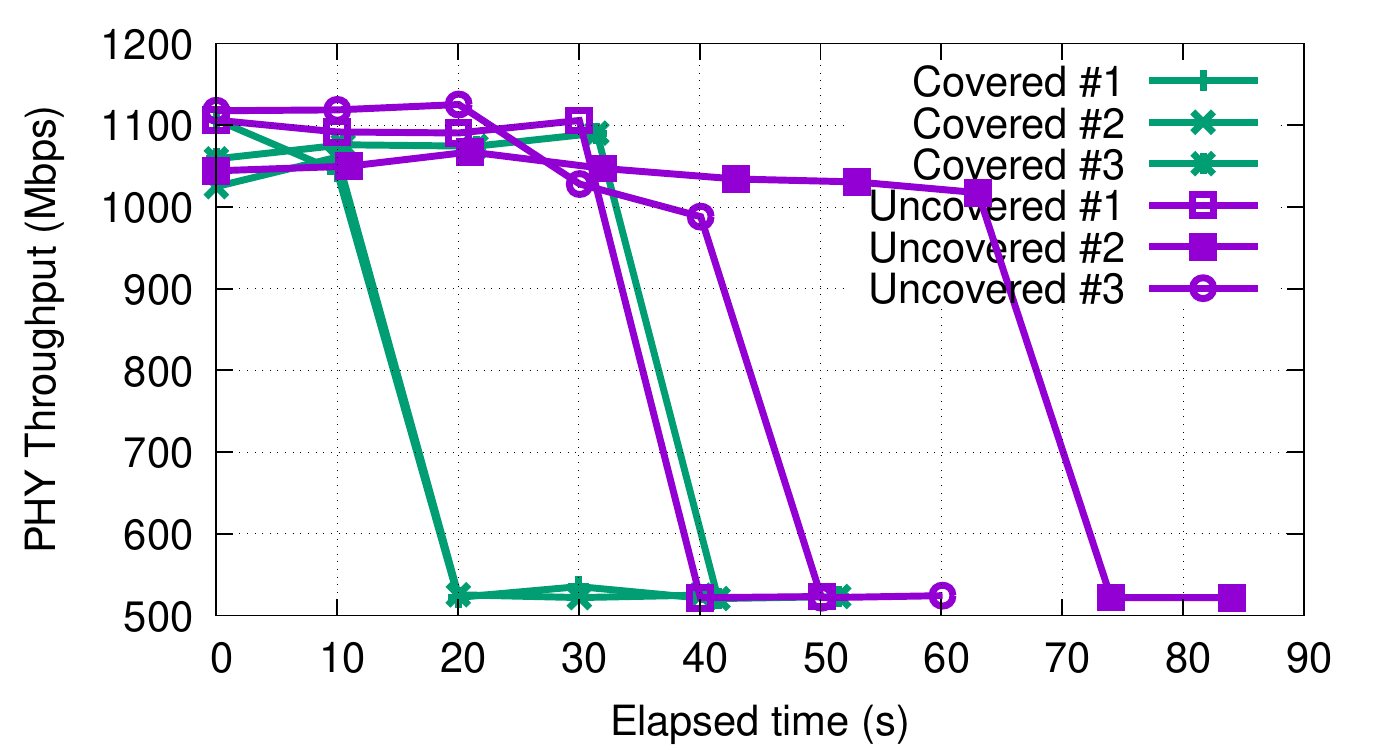}
        \caption{Throughput degradation between covered and uncovered phones.}
        \label{fig:cover_throughput}
    \end{subfigure}
    \begin{subfigure}{\linewidth}
        \includegraphics[width=1\linewidth]{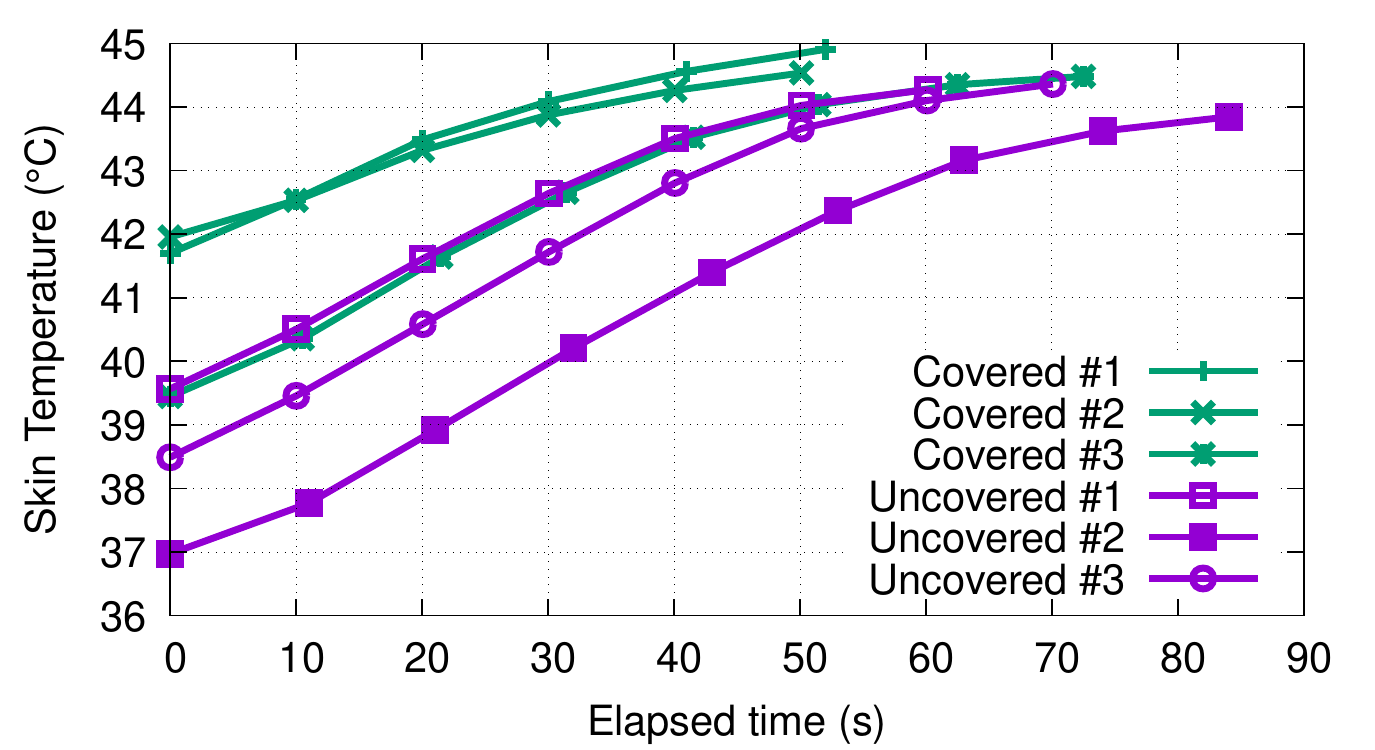}
        \caption{Skin temperature rise between covered and uncovered phones.}
        \label{fig:cover_temp}
    \end{subfigure}
    \caption{Phone cover experiments in Miami.}
    \label{fig:cover_exp}
    \vspace{-0.5cm}
 \end{figure}

Second, we investigate the impact of using a phone protective case on extending the 5G mmWave throughput. Using a standard commercially available case\footnote{\url{https://www.spigen.com/products/pixel-5-case-tough-armor}} to cover the phone, we compare throughput performance when the phone is with and without cover. Each experiment type is repeated 3 times and both types were ran in Miami. 
Fig.~\ref{fig:cover_throughput} shows the achievable PHY throughput for all six runs. Without a phone cover, the phone can sustain up to $60$~sec of a higher $\sim$$1$~Gbps 5G mmWave throughput using $4$ channels. With the cover, the phone can only sustain the higher rate up to $30$~sec.



The lower throughput performance of the covered phone can be explained by Fig.~\ref{fig:cover_temp}, which shows the corresponding skin temperature over all six runs. The covered phone breached the $43\degree$~C threshold at $20$~sec, compared to the uncovered phone at $40$~sec.
Hence, the phone cover restricts heat dissipation and causes a higher skin temperature. While further experiments with more variables (\eg phone model, case type, climate) are needed, this experiment has demonstrated that faster heat dissipation allows for longer utilization of the 5G mmWave network at full capacity.

\subsection{Thermal performance investigation using an IR camera}

In addition to extracting the skin temperature from the API, we also performed IR camera measurements in May 2022.
We set a FLIR One Pro LT IR camera up to mount stably $\sim$17.5~cm above a case-less Samsung S21+ phone with the Qualipoc measurement tool running\footnote{\url{https://www.rohde-schwarz.com/us/products/test-and-measurement/network-data-collection/qualipoc-android_63493-55430.html}}. The phone initiates BG DL traffic similar to prior experiments to capture PHY layer data (e.g., per channel SS-RSRP, SS-RSRQ, and throughput) and HTTP/application layer throughput.
We captured 4 $\times$ 10 minutes runs, with the phone connected to the 5G mmWave BS at Location 1 in Fig.~\ref{fig:Chicago_map} with a good signal condition ($\sim$90~dBm RSRP).
The ambient temperature at the time of the experiment was 30$\degree$~C.

\begin{figure}[!t]
    \centering
    \begin{subfigure}{0.45\linewidth}
        \includegraphics[width=1\linewidth]{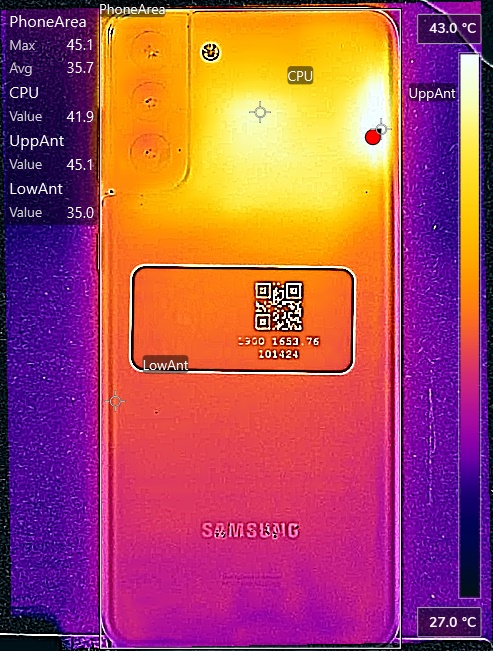}
        \caption{IR thermal capture on O1 (BS is to the right of phone).}
        \label{fig:flir_O1}
    \end{subfigure}
    \begin{subfigure}{0.45\linewidth}
        \includegraphics[width=1\linewidth]{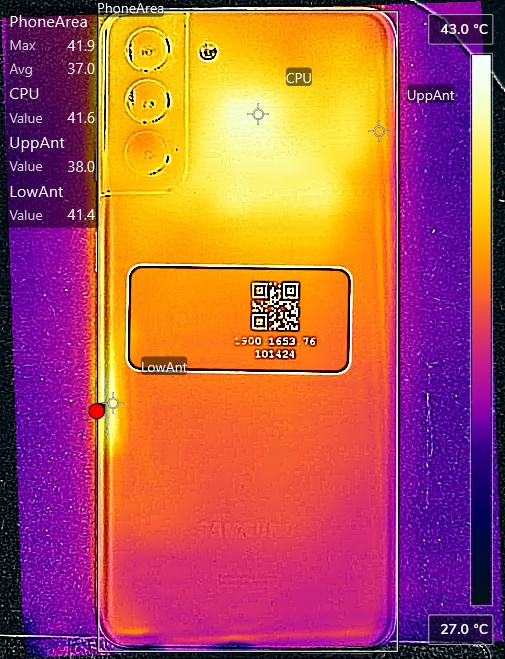}
        \caption{IR thermal capture on O2 (BS is to the left of phone).}
        \label{fig:flir_O2}
    \end{subfigure}
    \caption{IR thermal captures in orientations O1 and O2.}
    \label{fig:flir}
    \vspace{-0.5cm}
 \end{figure}
 
\begin{figure}[!t]
    \centering
    \begin{subfigure}{\linewidth}
        \includegraphics[width=1\linewidth]{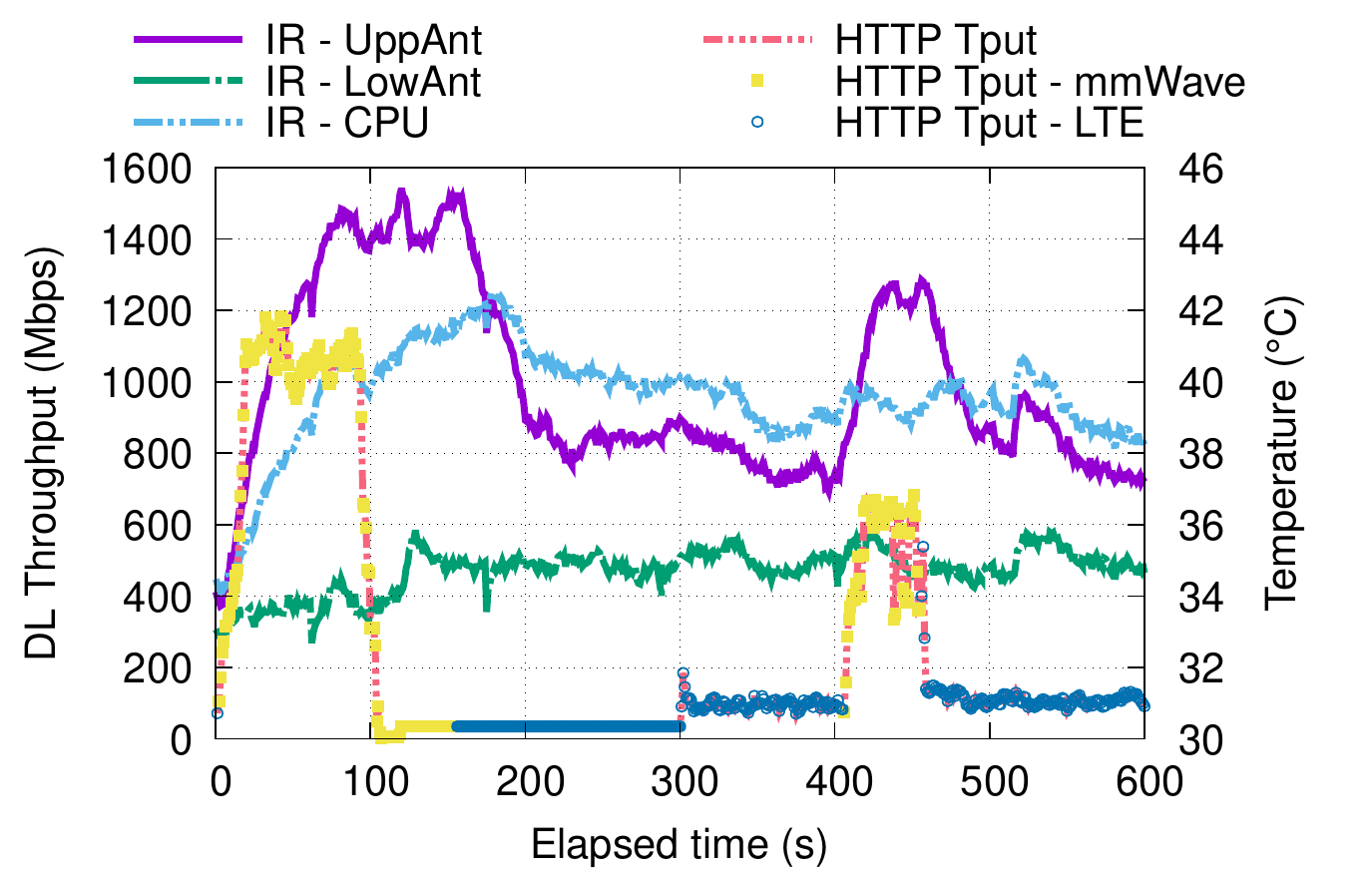}
        \caption{Throughput and temperatures vs. time in orientation O1.}
        \label{fig:time_tput_flir_O1}
    \end{subfigure}
    \begin{subfigure}{\linewidth}
        \includegraphics[width=1\linewidth]{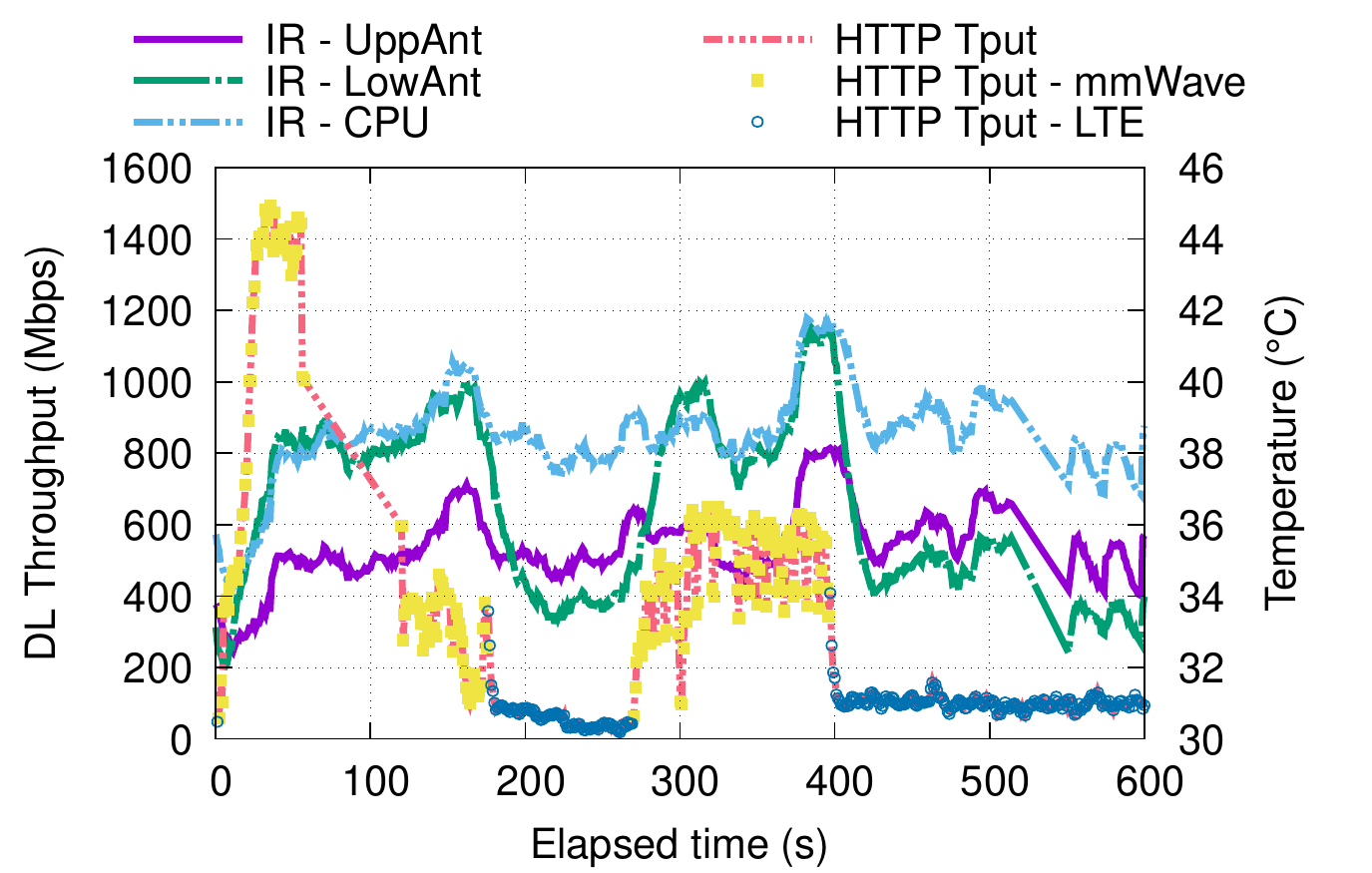}
        \caption{Throughput and temperatures vs. time in orientation O2.}
        \label{fig:time_tput_flir_O2}
    \end{subfigure}
    \caption{Throughput and temperatures vs. time.}
    \label{fig:time_tput_flir}
    \vspace{-0.5cm}
 \end{figure}

We identified three spots on the phone with a high likelihood of heating up during the mmWave experiment: CPU and modem area (CPU), upper mmWave antenna (UppAnt), and lower mmWave antenna (LowAnt).
x-y coordinates are defined for each spot, relative to the phone frame to ensure that the data is comparable between different experiment runs. 
Note that none of these temperature spots directly translate to the CPU and skin temperature value collected by the Android API since the temperature API is not directly accessible.


Fig.~\ref{fig:flir_O1} and Fig.~\ref{fig:flir_O2} show representative thermal images\footnote{Full video available at \url{https://youtu.be/fm29QwdbVW8}}, with two orientations, O1 and O2, w.r.t. the mmWave BS. In O1, the BS is located to the right of the phone, and both Fig.~\ref{fig:flir_O1} and Fig.~\ref{fig:time_tput_flir_O1} shows a higher temperature on the UppAnt spot which indicates higher activity on this antenna. 
Conversely, Fig.~\ref{fig:flir_O2} shows the O2 orientation where the BS is located to the left of the phone, and Fig.~\ref{fig:time_tput_flir_O2} shows higher LowAnt temperature when the phone is connected to mmWave.
In both orientations, the phone started with 4 mmWave channels then throttled to 1 channel at the 95~sec and 85~sec mark for O1 and O2, respectively. Subsequently, LTE handover occurred at the 154~sec and 174~sec mark for O1 and O2, respectively. We observed a better overall throughput performance on O2 compared to O1 on all 4 runs. This is due to two reasons: location of each antenna w.r.t. the CPU (O1 is closer), and which antenna is activated during the experiment. 

While we could not correlate the skin temperature to the spot temperatures, it is highly likely that the skin temperature sensor is located in the upper half of the phone, given that average temperature of the lower half was $\sim$35$\degree$~C on all runs. 
Additionally, we performed UL traffic experiments using a similar setup and observed that the phone was connected to 1 mmWave channel during the entire 10-minute run, even as the temperature of the UppAnt spot increased to a maximum of 42$\degree$~C.
This observation further supports our hypothesis, \ie fewer aggregated mmWave channels (and consequently, lower throughput) do not cause a significant rise in skin temperature,
even with the additional power consumption due to transmitting: {\bf transmitting} over 1 mmWave channel caused less rise in skin temperature compared to {\bf receiving} over 4 mmWave channels.

\section{Conclusions and Future Work}
\label{sec:conclusions}

We presented the first detailed measurements demonstrating the dramatic impact of device skin temperature on 5G mmWave throughput. The experiments were conducted in three different cities under various ambient conditions, with all results indicating that 5G mmWave sustained throughput is limited due to the rising skin temperature of the UE. We demonstrated a repeatable 3-step throughput profile which starts with a high rate above $1$~Gbps due to aggregating $4$ mmWave component carriers, which is downgraded to $1$ mmWave carrier before finally falling back to the baseline 4G/LTE system as the device skin temperature increases. Further, we have shown that the duration of sustained 5G mmWave throughput can be significantly increased by not using a phone covering or by using improved cooling mechanisms (e.g., an ice-pack for proof of concept purposes). Finally, IR imaging confirms that the mmWave antennas are a major contribution to the skin temperature rise. These results indicate that device skin temperature \textbf{should be considered} in scheduling and resource allocation algorithms so that the user does not experience a fluctuating throughput and the device does not heat up beyond the skin temperature limit of the phone. Our future work in this area will focus on
profiling the effects of CPU, modem, and mmWave antenna utilization on skin temperature, thermal throttling, and throughput.\footnote{All measurements collected for this paper will be made available on request.}

\bibliographystyle{IEEEtran}
\bibliography{IEEEabrv,references}

\end{document}